\documentclass[doublecol]{epl2}
\usepackage{graphicx}
\usepackage{colordvi}
\usepackage{amsmath}
\usepackage{amssymb}
\pacs{62.20.mm}{}
\pacs{46.15.-x}{}
\pacs{62.20.mt}{}
\title{Study of the branching instability using a phase field model of inplane crack propagation}
\shorttitle{Phase field study of crack branching}
\author{Herv\'e Henry}
\institute{PMC, Ecole Polytechnique, CNRS 91128 PALAISEAU FRANCE}
\abstract{
 In this study, the phase field model of crack propagation is used to study the
 dynamic branching instability in the case of inplane loading in two
 dimensions. Simulation results are in good agreement with theoretical
 predictions and experimental findings. Namely, the critical speed at which the
 instability starts is about $0.48 c_s$. They also show that a full 3D approach
 is needed to fully understand the branching instability. The finite
 interface effects are found to be neglectable in the large system size limit
 even though they are  stronger than the one expected from 
 a simple one dimensional calculation.} 
\begin{document}
\maketitle
\newpage
\section{Introduction}
 
 The study of crack propagation has gained a lot of interest in the physical
community during the recent years\cite{marderreview}. Among the many reasons
that have driven this interest is the difficulty to understand the mechanism
leading to the branching instability\cite{branch1,branch2}  that prevents a
crack from reaching its theoretical limiting speed: the Rayleigh wavespeed.
During the branching instability, a single straight propagating  crack separates
into two sub cracks. While, the instability takes place in the
process zone (the microscopic  region ahead of  the crack tip where the interatomic
bond breaking  leading to the crack propagation occurs), its effects (namely the
propagation of two sub cracks) are macroscopic. 

\begin{figure}
\includegraphics[width=0.22\textwidth]{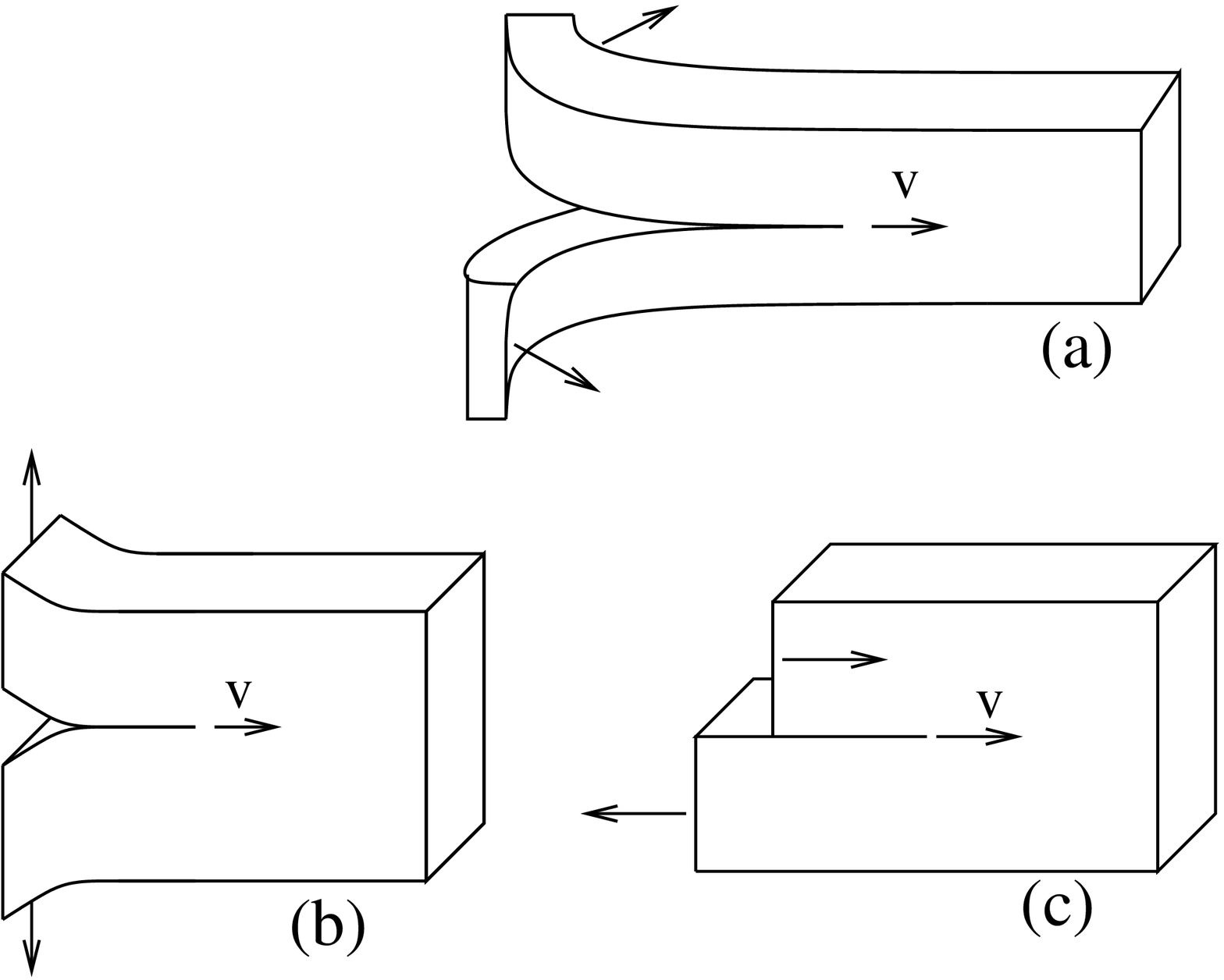}
\includegraphics[width=0.27\textwidth]{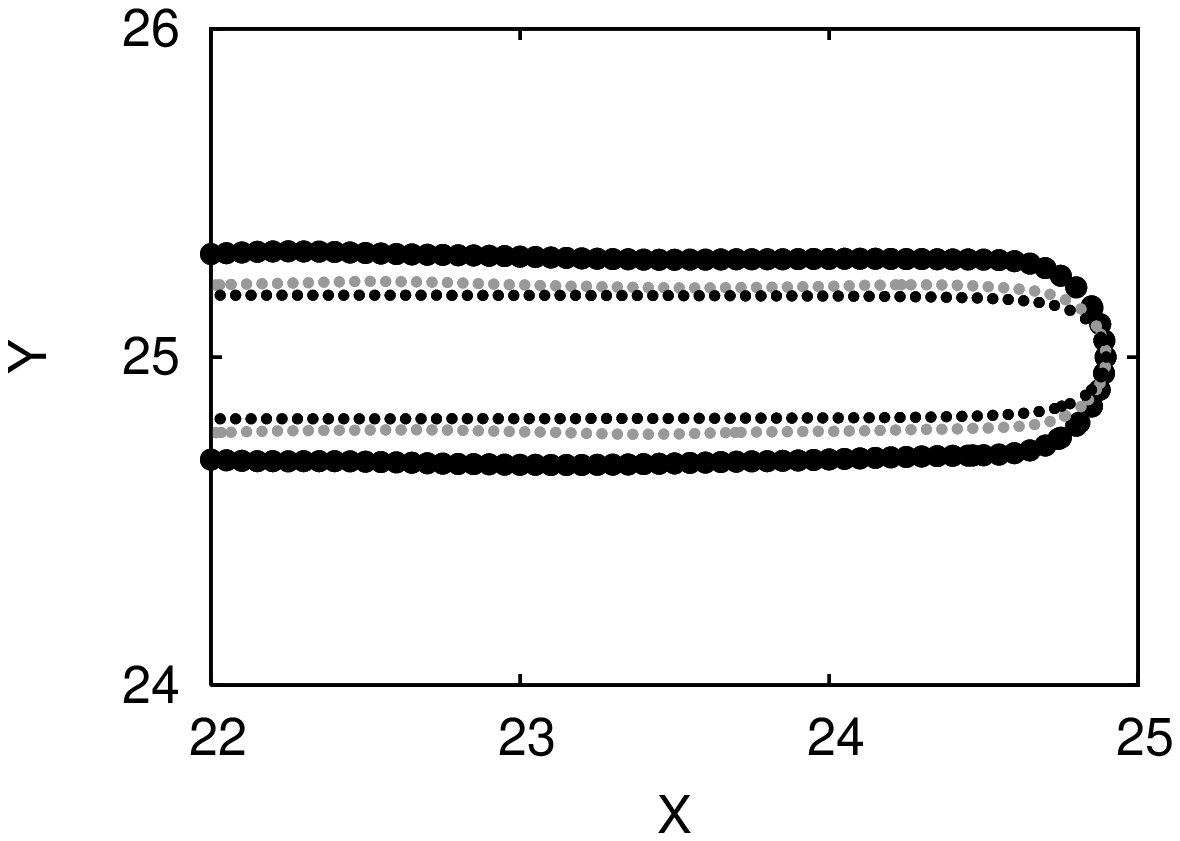}
\caption{\textbf{left} The three different modes of crack
propagation: (a) mode III, (b) mode I and (c) mode II. \textbf{right}Crack tip shape for different crack
speeds. Note the increase of the tip radius of curvature as the crack speed
increases ({\small $\bullet$} $v=0.12$, {\small \Gray{ $\bullet$}} $v=0.33$ and
{\large
$\bullet$} $v=0.47$ to compare with the branching
threshold of 0.52). For the higher speeds, the crack surface is not exactly
straight and presents some small irregularities.\label{fig_tips}}
\end{figure}

Hence a full description of the
branching instability needs to describe both the macroscopic and the microscopic
scales. The classical theory of crack propagation, describes well the
macroscopic scale where the linear elasticity theory is valid but describes the
crack propagation with  laws that postulate
the evolution  of the crack path is a function of the stress intensity factors
(number describing  the singularity at the crack tip, that correspond to the
three modes of crack propagation (see fig. \ref{fig_tips}). Such laws that do not
describe in any way the small scale of the process zone can not 
account for the onset of the branching instability unless one explicitly provides a criteria that
determines when a crack divides into two sub-cracks. Yet, they provide useful
results such as necessary conditions for branching\cite{addabranching,katzav2007}.

On the opposite, the use of molecular dynamics simulations can help
understanding the way a crack propagates and eventually divides into
branches,since it allows to compute what happens at the microscopic
scale.\cite{marder2004}
Nonetheless, it does not allow to reach 
neither long enough time scales nor large enough space scales to fully simulate  multiscale problems.   
 
 Therefore, in order to describe crack propagation an alternative approach may
be to use phenomenological models of crack propagation. Such models aim at
describing the behavior of the elastic material in the process zone with the
use of \textit{non-linear} elasticity. Typically, these models describe   the
process zone  as a region where the material softens gradually from a fully
intact state where the law of elasticity are verified to a fully broken state
where the crack cannot transmit any stress.

 A pioneering work was the introduction of the cohesive zone model where the
 crack line is prolongated by a softening segment\cite{barenblatt62,dugdale60}.
 More recently using idea from the phase transition theory, phase-field models
 of crack propagation were
 introduced\cite{Karmafrac,henrylevine2004,sethnaphasefield,igorphasefield,Spatschek2006,Pilipenko2007,Brener2003,corson2008}
 to describe crack propagation. They have  already proved they  can well
 describe the complex crack path in many cases\cite{henrylevine2004,Hakim2005}.
 In addition such models  have  been able to reproduce the feature of the crack
 branching in the case of mode III cracks\cite{Karmabranching,igorphasefield}. Here, a study of
 the branching instability using the phase-field model  is presented in the case
 of the inplane crack propagation (mode I and II). It must be emphasized here
 that while previous works have been devoted to the case of mode III crack
 propagation (paper tearing mode), this works is devoted to mode I crack
 propagation (pure tensile loading) which corresponds to usual experimental
 setups. 
 The convergence of the
 numerical results is discussed and  a comparison with available experimental
 data is presented.  Results show that the phase field approach is in good
 quantitative agreement with both theoretical predictions\cite{katzav2007} and
 experimental results, they also show that in order to fully understand three
 dimensional crack branching, three dimensional simulations are needed.  

\section{model and numerical implementation}

 The model used here has been introduced in the case of antiplane loading 
in\cite{Karmafrac} It has been shown to reproduce well the branching 
instability of cracks under antiplane loading\cite{Karmabranching}.  
The extension  of the model to crack propagation under inplane loading was 
presented in \cite{henrylevine2004} where it was shown to reproduce well 
the oscillating instability of cracks under bi-axial strain and also some
feature of the branching instability. This later model  was designed to comply
with  the principle of local symmetry and it has recently been used in the study
of quasistatic crack propagation under thermal load.\cite{corson2008} 

The principle  of the phase field approach is to introduce an additional
continuously (in space and time) varying variable $\phi$ that will describe the state of the elastic
material (e.g. $\phi=0$: broken and $\phi=1$:intact). It will evolve obeying
an evolution equation which couples $\phi$ with the  elastic fields. The equations of
motion of the elastic material are also coupled in an appropriate way with the
phase field. In this framework, the crack surface can be seen as an isocontour
of the phase field (e.g.  $\phi=0.5$).  Since $\phi$ is varying continuously,
the singularity at the crack tip is smeared out  and the  sharp tip is replaced
by a \textit{smooth} region where $\phi$ is varying rapidly from 1 to 0. This
region
may be considered as the process zone where failure occurs. This approach
differs from the cohesive zone approach\cite{barenblatt62,dugdale60} because
instead of smoothing the singularity by prolongating the crack line by a soft
interface line directed along the predicted \textit{ a priori } path of the
crack, it smooths the singularity at the crack tip by introducing a soft
region.  In this approach the crack faces are no longer lines but regions where
the elastic moduli of the material go from their nominal values to zero and
there is no need to know \textit{a priori } the crack path.   

First the model is described in more details.  The equation for the
displacement fields ($u_i$ where $i \mbox{ is in } x,\ y$)  writes as
in\cite{henrylevine2004}:
\begin{eqnarray}
\partial_{tt}u_i&=&-\frac{\delta E_{el}}{\delta u_i}\label{mecadot}\\ 
E_{El}&=&\int\int\int g(\phi)(\frac{\lambda}{2} (tr(\epsilon))^2+\mu(tr(\epsilon^2))
\end{eqnarray}
 where $\epsilon$ is the strain tensor and is, in the limit of linear
 elasticity, equal to $\epsilon_ij=(\partial_i u_j+\partial_j u_i)/2$. The
 function $g=4\phi^3-3\phi^4$ couples the phase field with the displacement
 fields and it has been chosen so that the stress transmitted through a crack
 vanishes in the large system limit (i.e. when the size of the system becomes
 much larger than the width of the crack interface introduced by the phase
 field.)\cite{Karmafrac}.  
The equation for the phase field writes:
\begin{eqnarray}
\beta \partial_t \phi&=&min(-\frac{\delta E_{\phi}}{\delta \phi},0)\label{eqphidot}\\
E_\phi&=&\int\int\int D(\mathbf{\nabla}\phi)^2 +\delta V(\phi)+g(\phi)(\mathcal{E}_\phi-\epsilon_c^2)
\label{eqenphi}\\
\mathcal{E}_\phi&=&\left\{\begin{array}{ll}
\frac{\lambda}{2}(tr \epsilon)^2+\mu tr \epsilon^2& \mbox{ if } tr \epsilon >0\\
\frac{\lambda-K_{\mathrm{Lam\acute{e}}}}{2}(tr \epsilon)^2+\mu tr \epsilon^2 &\mbox{ if } tr \epsilon <0
\end{array}\right.\label{eqephi}
\end{eqnarray}
 Equation (\ref{eqphidot}) insures that the phase field will not increase,
 which means that a crack can not heal. Equations (\ref{eqephi}, \ref{eqenphi}), are
 similar to the free energy used in \cite{Karmafrac}. The form of the coupling
 term in (eq. \ref{eqenphi}), indicates (if one neglects curvature effects at the
 tip) that the crack will propagate in a region where the value of
 $\mathcal{E}_\phi$ is higher than $\epsilon_c$. Contrarily to the model of
 \cite{Karmafrac}, here $\mathcal{E}_\phi$ is not the  local energy
 density. It takes into account the fact that the material is either under
 compression ($tr(\epsilon)<0$, $\mathcal{E}_\phi$ is the  contribution of
 shear to the local elastic energy density) or extension ($tr(\epsilon)>0$,
 $\mathcal{E}_\phi$ is the local elastic energy density) and that a crack
 should not propagate in a material under compression.  Therefore
 equations (\ref{eqephi}, \ref{eqenphi}) impose that in a material under
 compression, the favored state is intact material ($\phi=1$)
 
 This model depending on the imposed
 boundary conditions can  describe the propagation of a crack (or of many
 cracks) under arbitrary loading conditions. Model parameters  used here are
 $\epsilon_c(=1)$, $D(=0.25)$, $\beta(= 1,\ 2,\ 4 \mbox{ and } 8)$ and
 $\delta(=0.1)$. The fracture
 energy in the large system\cite{Karmafrac,henrylevine2004} limit is: 
  \begin{equation}
\Gamma=\sqrt{2D}\int_0^1\sqrt{\delta V(\phi)+\epsilon_c^2 (1-g(\phi))}
\end{equation}

 Here we have applied this model to the study of a dynamic crack propagating
 under mode I (tensile) loading: the top and bottom boundaries of the sheet are moved of
 $u_y(\pm W/2)=\pm\Delta_y/2$  and $u_x(x=0/L)=0$.  Parameters are $D=0.25$ so
 that the thickness of a crack ($w_\phi$)  at the onset is approximately 1 s.u.($D=1$ was
 also used without any significant change),
 $\lambda=1$, $\mu=1$, which implies that the shear wave speed is $c_s=1$ and
 the Rayleigh wave speed is $c_r=0.91$. The system is prepared at mechanical
 equilibrium with an initial pre-crack. 
Then the equations (\ref{mecadot},\ref{eqphidot}) are simulated on a regular grid
(with grid spacing $dx=0.1$) using a forward  Euler scheme. The discretization
scheme is chosen to keep the mechanical energy constant (if the phase field is
kept constant and so that the total energy decreases when the phase field
evolves.). In addition, checks were performed to insure that discretization
effects are neglectable: dividing the grid spacing  by 2 did not affect
results by more than 1 percent.

 Since the focus of this study is the stability/instability of straight cracks
 propagating at constant speed, simulations were  performed on a grid moving with
 the crack tip in order to let the transient accelerating regime disappear.
 The simulation grid had an aspect ratio ($L/W$) of 1. Simulations using larger
 and smaller aspect ratios (0.5, 2 and 4) showed a small change in quantitative
 results: namely an approximately  drop of the branching speed of 2 percents in
 the infinite aspect ratio limit for the three smallest width (12.5 25 and 50
 and 100) and $\beta=1$. On the contrary, changing the ratio $W/w_\phi$, where $w_\phi$
 is the interface thickness of the crack, did  affect quantitatively the
results. But, as expected\cite{Karmafrac}, results converge towards a limit when
the size of the sample $W$, compared to the width of the crack $w_\phi$,  goes to infinity 
(keeping the elastic energy stored in
the uncracked medium ($(\lambda/2+\mu)L(\Delta_y/L)^2$) per unit length and
other parameter constant). Therefore the fact that results  are affected 
when changing $W$ does not indicate that the branching is size dependent. It
is indicative of 
the effects of the finite width of the crack in the phase field
model (It is not due to the fact that the full stress relief only occurs in the
infinite $W$ limit, which is a 1D effect. It is due to the fact that a finite
distance exists between the edges of the crack).

\section{Straight crack}

 \begin{figure}
\includegraphics[width=0.23\textwidth]{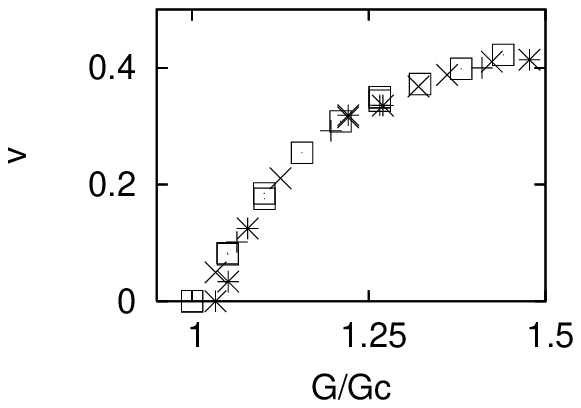}
\includegraphics[width=0.23\textwidth]{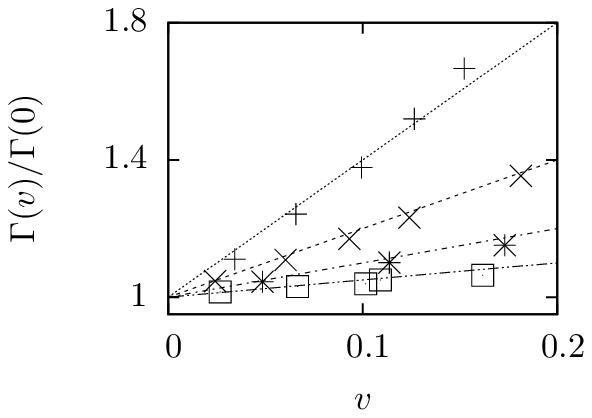}
\caption{\textbf{left}Crack speeds as a function of the available elastic energy
($\propto\Delta_y^2/W$) normalized by the  crack energy (derived from the
model equations as in \cite{Karmafrac}) for $\beta=1$ and different system sizes.
 One should note that the curves collapse extremely well on a master curve
and that, for the sake of simplicity, only the regime of straight crack
propagation is shown.\label{fig_steady_speeds}\textbf{right} Fracture energy
normalized by the fracture energy at zero velocity computed using eq.\ref{eq_freund90} as a
function of crack speed for different values of $\beta$ ($\beta=$1
($\square$),2 (*),4 (X) 8 (+)). The lines correspond to
the law eq. \ref{eq_dissipation} and fit well the numerical datas. For higher
values of $v$ one can observe a deviation from the linear law that corresponds
to the onset of the branching instability.} 
\end{figure}
First some results on the propagation of a stably propagating single crack are
presented. Using the model parameters, the fracture energy is $2\Gamma=0.971$.
According to the Griffith criterion, one expects that a 
crack will begin to propagate when the elastic energy stored per unit length is
larger than the energy needed to create two crack interfaces, that is for
$\Delta_y=\sqrt{(\Gamma W)/(\lambda/2+\mu)}$. Simulations results show that this
prediction is extremely well verified, with an error smaller than a percent, for
the smallest simulation boxes used here (W=25). For bigger systems (W=50, 100
and 200), the error is always at most of the same order of magnitude. Hence the
model used here is in very good agreement with the Griffith criterion.

 When the value of $\Delta_y$ is increased, the speed of the
 straight crack $v$ is expected to obey the law:
 \begin{equation}
 G A_I(v)=\Gamma(v)\label{eq_freund90}
 \end{equation}
 where $\Gamma(v)$ is the mechanical energy that is either disspated or
 converted into surface energy  when the fracture advances of one space 
 unit at speed $v$, G is the
 mechanical energy available far ahead from the crack tip and $A_I(v)$ is a
 universal function that depends solely on the elastic properties of the
 material (for further details see ref.\cite{Freund} sections  5.3 and 5.4).
 Simulation results using the phase field model and  eq. \ref{eq_freund90} show that 
 \begin{equation}
 \Gamma(v)=\Gamma(0)(1+a_\beta\beta v),
 \label{eq_dissipation}
 \end{equation} 
 with $a_\beta \approx 0.5$ a constant, for   
 values of $v$ below the branching threshold, and indicate that for $\beta\to
 0$, the fracture energy is independant of the crack speed (zero dissipation
 limit). It must be noted here that for or a given model
 parameter set, the relation between the steady crack speed  $v$ and
 the available mechanical energy ahead of the crack tip is independant
 of the system size (see fig. \ref{fig_steady_speeds}). 
 
 At this point, before turning to the study of branching instability, it is
 worth to summarize the results obtained here. First the phase field model, as
 already shown in \cite{henrylevine2004} obeys the Griffith criterion. In
 addition, the relation between the model parameter $\beta$ and the speed
 dependance of the fracture energy has been computed (eq.\ref{eq_dissipation}).
 The later result together with theoretical prediction in \cite{katzav2007}
 which show that the branchinsg speed $v_c$ behaves like
 $a+b(\Gamma(v_c)-\Gamma(0))/\Gamma(0)$ ,
 allows to predict  that the branching speed in the phase field model
 must behave linearly with $\beta$ (taking into account the fact that in
 \cite{katzav2007}, only a necessary condition for branching was derived while
 here we focus on the onset of branching).
 
    We now turn to the main purpose of this work: the study of the branching instability. 
\section{branching instability}
\begin{figure}
\includegraphics[width=0.5\textwidth,height=3.cm]{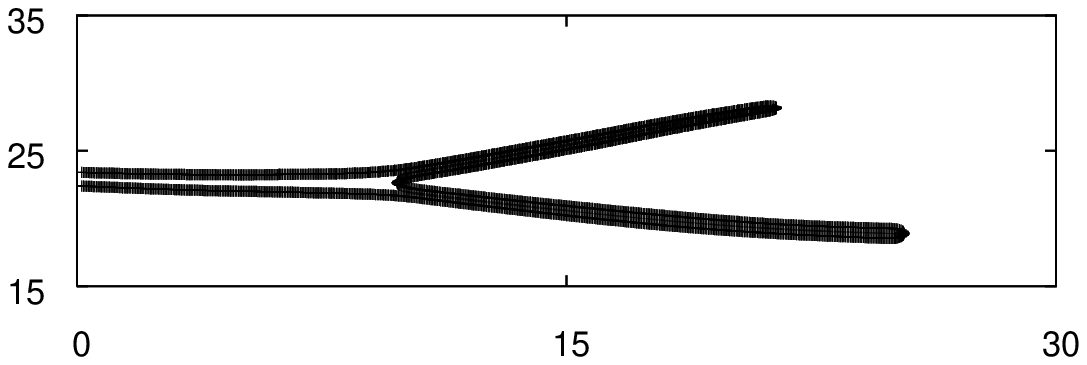}
\includegraphics[width=0.5\textwidth,height=3.cm]{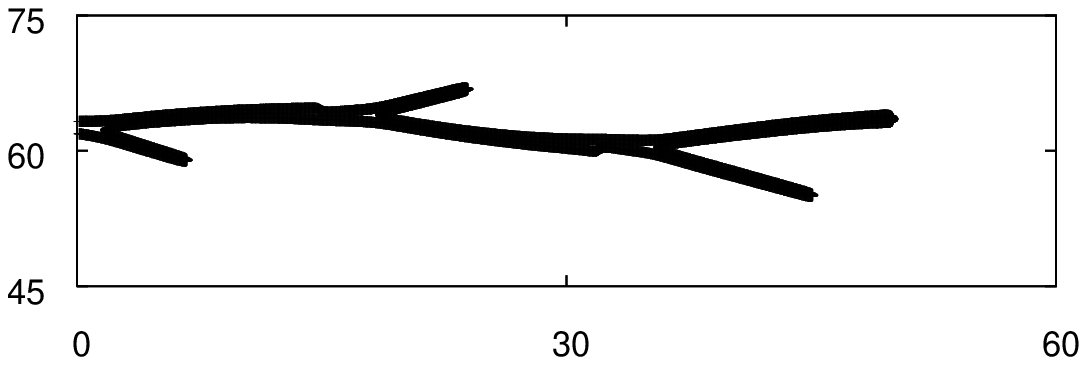}

\caption{Typical crack paths  during the branching instability (the crack is
propagating from left to right). The top plot
shows a single branching  event.  The bottom plot shows a fast
crack with multiple macroscopic branches appearing.
 In the top plot, the middle of the elastic plate
is at $y=25$, and in the bottom  it is at $y=50$. Parameter values are:
$\beta=1$ and $G/G_c=2.5$\label{fig_branchingcracks}}
\end{figure}
\begin{figure}
\includegraphics[width=0.45\textwidth]{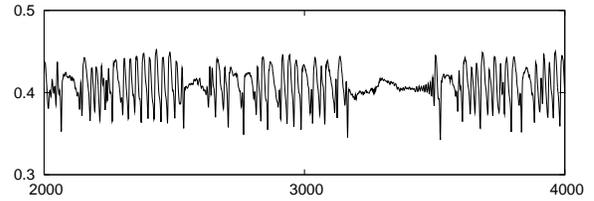}
\caption{\label{tracelongue}Speed of the crack as a function of time. One can
see the intermittent behavior of the crack speed. The burst of crack speed
oscillation correspond to branching events while the regions where the speed is
varying slowly correspond to a situation where the crack has divided into two
branches that propagate together for a while before one of them takes
over. $\beta=8$.}
\end{figure}

 When the value of $\Delta_y$ is further increased, the crack tip becomes
 unstable and the crack divides into two  cracks. As shown in
 \cite{Karmabranching}, this instability occurs through \textit{tip blunting}.
 As the crack speed increases, the radius of curvature of the crack tip
 increases  and when the speed is larger than a critical value, the crack tip
 splits into two branches that propagate (see fig.
 \ref{fig_branchingcracks}, where one can see that the crack tip is still
 much smaller than the system width). 
  One of those two branches can be
 faster than the other and through \textit{screening} of stress eventually
 prevents the propagation of the other branch. Then if strain is sufficient, the
 scenario repeats and one can see a succession of branching events giving birth
 to two cracks, the selection of one of them its acceleration and separation
 into two cracks... This regime of sustained disordered branching events can
 only take place if strain is sufficient and corresponds to the macro-branching
 instability (see fig. \ref{fig_branchingcracks}). 
 Its signature when considering  the crack surface is the presence
 of successive branches of various sizes. In addition to this qualitative
 description, one notices a sharp increase in the variance of the crack speed
 which is similar to the one observed in experiments despite some
 discrepancies (The variance of the crack speed as a function of the
 mean crack speed increases via a
 discontinuous jump in 
 2D simulations while in experiment it increases  linearly above threshold)
 which could be attributed to the 3D nature of experiments. This transition is easily noticed when considering the velocity
 profile (Here crack velocity is the time derivative of the position of the most
 advanced point of the crack: the iso contour $\phi=05$). 
 Instead of small oscillations one can see sharp decreases of the
 crack speed followed by a progressive  acceleration of the crack and again a
 sharp decrease (see fig. \ref{tracelongue}). This behavior of the crack speed
 is typical of a branching event and is in  good agreement with the scenario
 presented in \cite{katzav2007}, with numerical observations of
 \cite{Karmabranching} and with experimental results. 

\begin{figure}
\includegraphics[width=0.25\textwidth]{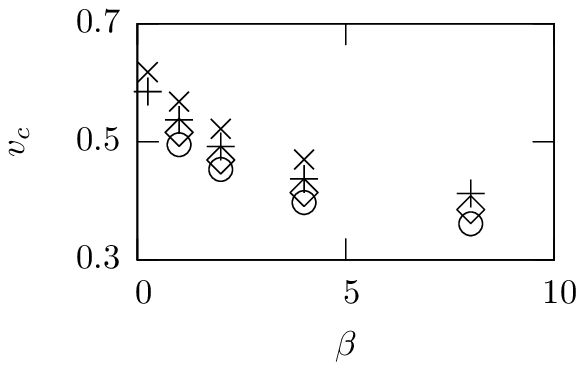}~
\includegraphics[width=0.25\textwidth]{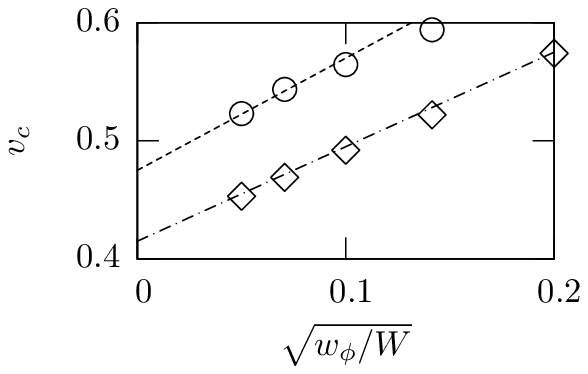}
\caption{\label{fig_speed_branching}\textbf{(a)} critical speed for which the
straight crack is no longer 
stable as a function of $\beta$ for different value of $W/{w_\phi}$, the ratio of the
system size (W) on the width of the phase field crack (${w_\phi}$. One should note that as 
$\beta$ goes  toward zero (zero dissipation
limit), the critical speed converges toward a value that depends on the system
size and that contrarily to what is observed in fig.\ref{fig_steady_speeds}
the curves do not collapse on a master curve. ($\times$, $+$, $triangles$ and circles correspond
respectively to sizes of 25, 50, 100 and 200). \textbf{(b)} Branching speed in
the zero $\beta$ limit (circles) and for $\beta=2$ (diamonds) as a function of
the system size $W$ relative to the crack width ($w_\phi$). 
The lines are used as a guide to the eye. The fact that
the convergence of the speed is in  $\sqrt{w_\phi/W}$ may be attributed to
the effects of the finite width of the crack interface on the singular
behavior at the crack tip.}
\end{figure}

 The critical speed depends on the model parameters (here:
$\beta$\footnote{Additional simulations have shown that dividing $\delta$ by 10
did not affect results and simulations of\cite{Karmabranching} in the case of
mode III cracks show that $\delta\to 0.$ is the correct limit in the sense that
it allows to retrieve theoretical predictions}) and on the system size (see
fig. \ref{fig_speed_branching}). While
the dependence on $\beta$ is expected since $\beta$ measures the energy
dissipation at the crack tip, the dependence on $W/w_\phi$ is not the result of
any physical phenomena and comes from the way the phase field models deals with
the singularity at the crack tip.

Indeed, it does not allow to consider an
infinitely sharp crack. It only allows to reproduce a crack of finite width
$w_\phi$
and one should retrieve the results of an infinitely sharp crack in the limit
$w_\phi/W \to 0$. Simulations results show that this is actually the case. In the
limit $w_\phi/W \to 0$, the value of the critical speed at which the macroscopic
branching instability occurs for a given value of $\beta$ converges toward a
finite limit while the qualitative behavior of the system remains unchanged. As
mentioned earlier, this limit is a function of $\beta$ which measures the
dissipation in the material. As a result, in the limit $\beta \to 0$, the phase
field model reproduces the behavior of a perfectly brittle material. Simulations
using different values of $\beta$ show that the branching speed scales linearly
with $\beta$ and converges toward a finite value when $\beta \to 0$ as one
would expect from \cite{katzav2007}  and the results of the study of steady
crack propagation.

  Then in the limit $\beta\to 0,\ (w_\phi/W)\to 0$, one gets the value of a critical
macroscopic branching speed for a perfectly brittle material without any
dissipation of approximately 0.48. This value is in fairly good agreement (a $15\%$
discrepancy) with theoretical results of\cite{katzav2007} which predicts a
branching speed of 0.42 (with  parameters used here). The discrepancy can be
easily attributed to the fact that in \cite{katzav2007}, the critical speed that
was computed, was the speed at which branching can occur, i.e. a necessary
condition, while here the computed speed is the one at which a straight crack is
no longer stable. And to reinforce this, it is important to mention that during
simulations branching events were observed at instantaneous  speeds lower than
the the measured critical velocity. 

When comparing, the results obtained here with the
experiments\cite{Sharon1995,Sharon1996,Sharon1996a}, one must first note that
the three dimensional character of the branching instability cannot be
reproduced by our simulations and that in PMMA, according to \cite{Sharon1996a}
(figs. 19 and 17), the branching instability can be described as 2 dimensional
for average  crack speeds much higher ($\approx 0.58 c_R$) than the critical
	speed for the 3D branching instability ($0.34c_R$).  In addition,
	dissipative effects are present in PMMA and are not well quantified
	above  the first critical speed ($0.34 c_R$). Then, it is clear that
	our simulations can not be compared quantitatively with the
	experimental results. Nonetheless, some qualitative comparison are
	possible. First of all, the computed branching angle defined as the
	angle between the secondary crack and the main direction of
	propagation of the crack is in  good
	agreement with experimental results. It is on average equal to $27^o$
	(with some visible variations), 
	which is close  to  the experimental value of  $30^o$.  Interestingly
	enough, if one defines the branching angle as the angle
	between the two branches the computed values is then around $45^o$
	which is in rather good agreement with theoretical predictions of
	$54^o$\cite{addabranching,katzav2007}. The
	discrepancy in the computed values of angles  
	may come from the fact that the phase field method  does
	not allow to compute the branching angle at the onset of the branches
	leading to a small under-estimation of its value. Indeed, the branching
	point is  defined up to a few interface thicknesses  and the behavior
	of the branches can only be described a few interface thicknesses away
	from the branching point. These shortcomings of the phase field
	approach do not allow to check the asymptotic behavior of branches
	close to the onset of branching. Simulations in larger systems are
	needed to overcome the finite interface limitation.
 
\begin{figure}
\includegraphics[width=0.5\textwidth]{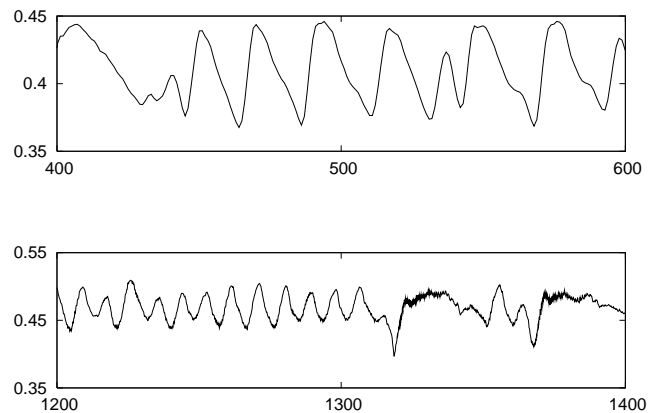}
\caption{\label{speedtrace} Typical curves of the crack speed as a function of time  during
repetitive branching events for two different values of $\beta$ (top: 8  and
bottom: 4). One can see that the typical interval between maxima's of the
$\beta=8$ curve are approximately two times bigger than in the $\beta=4$ curve.}
\end{figure}

From a \textit{statistical} point of view, the crack speed fluctuations power spectrum in the case where no dissipation occur in the bulk did  not exhibit any clear frequency.  
The same behavior was observed in the case where a small dissipation was added
in the bulk. 
  Nonetheless, during the repetitive branching burst, the frequency
of branching events was found to be fairly constant for given model parameter
values. Namely its Independence on the system size, on the crack speed could
not be determined. Nonetheless, a clear dependence on the value of $\beta$ was
found: the inter-branch  interval is approximately proportional to $\beta$ as
can be observed in fig. \ref{speedtrace}. This behavior is true for small
frequencies (i.e. for $\beta>2$). But for small values of $\beta$ (lower than 2), 
one can see that the inter-branch
intervals ceases to  decrease and saturate at a given value. This saturation
may be attributed  to the finite  thickness
of the interface (there exists a minimal interbranch distance which
can be estimated to be equal to a few interfaces thicknesses. Here, assuming a
crack speed of around $0.5c_s$, and considering the crack width of 1. su.,
the highest frequency should be a fraction of 0.5.\footnote{It cannot be
more than 0.5 and it is reasonable to assume that one needs at least a few crack
width between branches.}).Hence as $\beta$ is decreased, phase field
simulations show an increase of the typical frequency until it reaches the
saturation frequency.  

\section{Discussion}
 
 Results of simulations presented here show that the phase field approach is
 useful to understand fracture as a pattern formation mechanism. Nonetheless,
 the fact that crack propagation is due to the $1/\sqrt{r}$ singularity at its
 tip, finite size effects due to the phase field approach turn out to be much
 stronger than one would expect from the 1D calculation of \cite{Karmafrac}. 
 This slow convergence does not prevent the model to retrieve quantitatively 
 good results in the large system
 size and the use of better simulation methods (such as Finite elements or
 adaptive remeshing) should allow to consider systems where finite size
 effects are neglectable. Hence, the phase field model is able the reproduce
 correctly the branching instability in the case of 2D crack propagation,
 predicting a critical speed of $0.48c_s$, which is $15\%$ larger than the
 necessary condition derived in \cite{katzav2007}. Since
 the branching instability is essentially a 3D process\cite{Sharon1996a},
 comparison with experiments is only indicating that the phase field model 
  can reproduce the main features
 of the branching instability. But results and comparison with experiments
 indicate that  full 3D simulations are needed to
 be able to understand 
 the branching instability. One of the expected results of this is to be able
 to understand what are the parameter that govern the formation of fracture
 patterns such as the parabolic markings  shown in \cite{Sharon1996a}.

\acknowledgments
I wish to thank Mokhtar Adda-Bedia, Alain Karma, Eran Sharon and Jay Fineberg
for fruitful discussions during this work.

\end{document}